 \documentstyle[epsf,times]{mn} 

\font\twelvei = cmmi10 scaled\magstep1
       \font\teni = cmmi10 
\font\mbf = cmmib10 scaled\magstep1
       \font\mbfs = cmmib10 \font\mbfss = cmmib10 scaled 833
\font\msybf = cmbsy10 scaled\magstep1
       \font\msybfs = cmbsy10 \font\msybfss = cmbsy10 scaled 833
 
\font\msybf = cmbsy10 scaled\magstep1
       \font\msybfs = cmbsy10 \font\msybfss = cmbsy10 scaled 833
\def\etal{{\rm et al.\ }}
\def\eg{{\rm e.g.\ }}
\def\ie{{\rm i.e.\ }}
\def\lsim{\mathrel{  
        \raise0.3ex\hbox{$<$}\kern-0.75em{\lower0.65ex\hbox{$\sim$}}}}
\def\gsim{\mathrel{
        \raise0.3ex\hbox{$>$}\kern-0.75em{\lower0.65ex\hbox{$\sim$}}}}
\def\kms{\mbox{\rm\,km\,s$^{-1}$}}      
\def\SSS{\scriptscriptstyle}

\def\sun{\odot}
\def\kpar{\kappa_{\SSS\|}} 
 
\begin{document}
\title[Origin of Highest Energy Cosmic Rays]
{Contributions to the Cosmic Ray Flux above the Ankle:
\\       Clusters of Galaxies}
 
\author[H.Kang, J.P.Rachen and P.L.Biermann]
{Hyesung Kang$^{1,2}$,
J\"org P.~Rachen$^{3,4}$ and Peter L.~Biermann$^{3,5}$ \\
$^1$Department of Earth Sciences, Pusan National University,
    Pusan 609-735, Korea\\
$^2$kang@astrophys.es.pusan.ac.kr\\
$^3$Max-Planck-Institute for Radioastronomy, D-53010 Bonn, Germany\\
$^4$jrachen@mpifr-bonn.mpg.de\\
$^5$plbiermann@mpifr-bonn.mpg.de}
 
\maketitle
 
\begin{abstract}
 
Motivated by the suggestion of Kang, Ryu \& Jones (1996) that particles can
be accelerated to high energies via diffusive shock acceleration process at
the accretion shocks formed by the infalling flow toward the clusters of
galaxies, we have calculated the expected particle flux from a cosmological
ensemble of clusters.  We use the observed temperature distribution of
local clusters and assume a simple power-law evolutionary model for the
comoving density of the clusters.  The shock parameters such as the shock
radius and velocity are deduced from the ICM temperature using the
self-similar solutions for secondary infall onto the clusters.  The
magnetic field strength is assume to be in equipartition with the postshock
thermal energy behind the accretion shock.  We also assume that the
injected energy spectrum is a power-law with the exponential cutoff at the
maximum energy which is calculated from the condition that the energy gain
rate for diffusive shock acceleration is balanced by the loss rate due to
the interactions with the cosmic background radiation. In contrast to the
earlier paper we have adopted here the description of the cosmic ray
diffusion by Jokipii (1987) which leads to considerably higher particle
energies. Finally the injected particle spectrum at the clusters is
integrated over the cosmological distance to earth by considering the
energy loss due to the interactions with the cosmic background radiation.
Our calculations show that the expected spectrum of high-energy protons
from the cosmological ensemble of the cluster accretion shocks could match
well the observed cosmic ray spectrum near $10^{19}$eV with reasonable
parameters and models if about $10^{-4}$ of the infalling kinetic energy
can be injected into the intergalactic space as the high energy particles.
 
\end{abstract}
 
 
\section{Introduction}
 
It is widely believed that diffusive shock acceleration is the mechanism
from which cosmic rays get their energy. There are various models to
account for the origin of cosmic rays below about $3{\times}10^{18}\,{\rm
eV} = 3\,$EeV.  At low energies, up to about $10^{14}$ eV, supernova
explosions into the interstellar medium give a reasonable and successful
explanation for the data for protons.  At higher energies, several models
have been proposed, most notably a galactic wind termination shock (Jokipii
\& Morfill 1987), and multiple shocks in an ensemble of OB superbubbles and
young supernova remnants (Axford 1992).  A comprehensive, albeit tentative,
theory has been proposed by Biermann (1993 and later papers) that explains
the cosmic ray spectrum, with its chemical abundances and the knee feature
as resulting from a combination of supernova explosions into the
interstellar medium, and supernova explosions into strong stellar winds of
progenitor Wolf-Rayet stars (Biermann 1993; Biermann \& Cassinelli 1993;
Biermann 1995). Above the ``ankle'' at about 3\,EeV (the ultra-high energy
cosmic ray regime, called UHECR hereafter), a simultaneous change in
spectrum and composition of the cosmic ray spectrum (Bird \etal 1994)
suggests a change of origin. As argued already by Cocconi (1956) particles
of higher energy need to come from outside our galaxy due to the very large
Larmor radius, and therefore need to be accelerated in extragalactic
sources.  Here we concentrate on these high energy particles that almost
certainly come from extragalactic sources.
 
It is well known that the extragalactic cosmic ray spectrum must show the
Greisen-Zatsepin-Kuzmin (GZK) cutoff at about 60\,EeV due to interactions
with the cosmic background radiation (CBR), regardless whether protons,
heavy nuclei or photons are considered as the energetic particles
(Greisen 1966; Zatsepin \& Kuzmin 1966). Pair
production and the cosmological evolution of radiation backgrounds set
limits for the distance of cosmic rays even at lower energies, thus any
extragalactic model of cosmic ray origin must propose at least some
sources in our cosmological neighborhood which can account for the highest
energies. In an expansion of an earlier proposition by Biermann \&
Strittmatter (1987) which predicted maximum proton energies in active
galaxies of $10^{21}$ eV, Rachen \& Biermann (1993) developed a model to
accelerate cosmic rays up to a few 100\,EeV at strong shocks at the end of
extended jets in powerful radio galaxies (FR-II radio galaxies, Fanaroff \&
Riley 1974). It has been shown that even for a moderate proton content in
the jets this model can generally account for the cosmic ray flux above the
ankle, and is consistent with air shower data suggesting a takeover from
heavy to light nuclei in this energy range (Rachen, Stanev \& Biermann
1993).  It can be shown that this model can be extended to a maximum
particle energy in the source of about $4{\times}10^{21}$ eV (Biermann
1996).  One prediction of this particular model has been tested
successfully, and that is the expected correlation of arrival directions of
very high energy cosmic rays with the large scale distribution of radio
galaxies, in the supergalactic plane (Stanev \etal 1995; Hayashida \etal
1996).

However, the uncertain proton content of the jets, the energy limitation
for cosmic rays due to the relatively small acceleration region in hot
spots, the finite life time of hot spots, and the large distance of the
closest well known FR-II radio galaxy set strong constraints on the
predictability of the model, and leaves room for other contributions to the
highest energy cosmic rays. In particular, after the detection of a
320\,EeV air shower by Fly's Eye (Bird \etal 1994), and another 200\,EeV
event by the AGASA ground array (Hayashida \etal 1994), the various
acceleration models have been critically reviewed (\eg Elbert \& Sommers
1995; Biermann 1996) and new models have been suggested.  Most of them are
based on astrophysical objects whose physical properties are under
controversial discussion, as the decay of topological defects
(Bhattacharjee 1991; Sigl, Schramm \& Bhattacharjee 1994; Protheroe \&
Johnson 1996; Protheroe \& Stanev 1996), or rapid acceleration in Gamma Ray
Burst sources (Milgrom \& Usov 1995; Vietri 1995; Waxman 1995), but the
poor statistics of events above 100\,EeV allows a variety of other
explanations, including the earlier proposal of radio galaxy origin (Rachen
1995; Stanev \etal 1995; Biermann 1995, 1996).
 
Recent work shows that there could be sites for shock acceleration in
extragalactic space alternative to the strongly constrained acceleration in
radio galaxy hot spots. According to hydrodynamic simulations of large
scale structure formation (\eg Kang \etal 1994a; Cen \& Ostriker 1994),
accretion shocks are formed in the baryonic component around non-linear
structures collapsed from the primordial density inhomogeneities as a
result of gravitational instability.  Those structures can be identified as
pancake-like supergalactic planes, still denser filaments, and clusters of
galaxies which form at intersections of pancakes, in any variants of the
many cosmological models.  They are surrounded by the hot gas heated by the
accretion shocks and the particles can be accelerated to very high energies
at these shocks via first order Fermi process.  Kang, Jones \& Ryu (1995)
and Kang, Ryu \& Jones (1996, KRJ96 hereafter) suggested that the accretion
shocks around the clusters of galaxies could be as fast as 1000--3000\,\kms
and so could be good acceleration sites for the UHECRs up to several
10\,EeV, provided there is a turbulent magnetic field so that the diffusion
is in the Bohm limit, and if the magnetic field around the clusters is
order of microgauss. We note here the maximum energy could be shifted to
super-GZK values, if the field geometry is close to being
quasi-perpendicular (Jokipii 1987).  The significance of the accretion
shocks is that they are the largest and longest lived shocks in the
universe, so that they naturally get past the primary factors normally
limiting Fermi acceleration to such high energies.  Especially the
accretion shocks formed around the clusters of galaxies which have the
deepest gravitational potential well are the strongest and thus could
accelerate the particles to the maximum possible energy.  Independently
Norman, Melrose \& Achterberg (1995) showed with the help of a similar
argument that the shocks associated with the large-scale structure
formation could accelerate the protons up to $E_{\rm max}=50\,$EeV if there
is a primordial field of 1--10 nanogauss, or if microgauss field can be
self-generated in shocks.  
 
In the present study we have estimated the contribution of the CR protons
from an ensemble of the cluster accretion shocks distributed in the
universe by adopting some simple models for the accretion flows onto
clusters, the strength, geometry, degree of irregularities of the magnetic
field near the cluster accretion shocks, and the cosmological evolution of
the cluster distribution.  The details of the models will be given in \S 2.
In \S 3 the estimated CR proton spectrum has been compared with actual
observations in order to see if the proposed origin could explain the
energy spectrum of UHECRs with reasonable physical parameters and models.
We discuss the implication of these results and models also in \S 3.
 
We adopt for the following $\Omega_0 = 1$ and write the Hubble constant as
$H_0 = 100 \, h \, \rm km/s/Mpc$; in numerical calculations we use
$h=0.75$.
 
\section{Models}
 
\begin{figure*}
\epsfysize=3.7in\epsfbox[56 510 555 780]{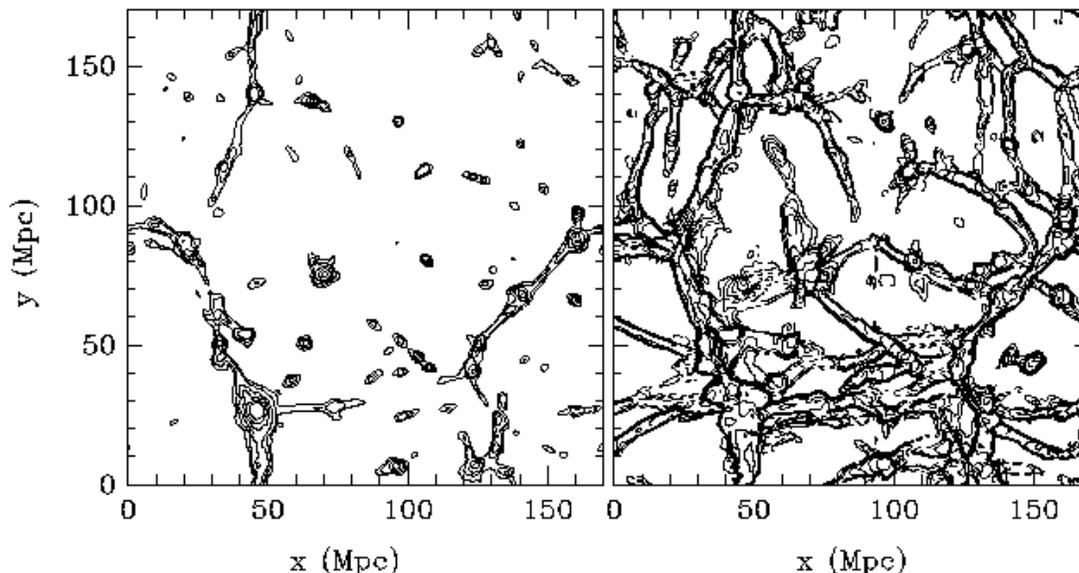}
\caption{A representative slice cut of a simulated universe in a standard
cold dark matter model.  The left panel shows the X-ray luminosity
distribution while the right panels shows the gas temperature distribution.
See text for the details.}
\end{figure*}
 
\subsection{Accretion Shocks around Large Scale Structures}
 
It is generally accepted that both galaxy distribution in the observed
Universe and matter distribution in numerically simulated universe in most
of generic cosmological scenarios show sheetlike and filamentary structures
on large scale (\eg de Lapparent, Geller \& Huchra 1991; Melott 1987; White
\etal 1987).  Numerical simulations based on even a hierarchical clustering
model such as variants of CDM models also indicate that the dominant
nonlinear structure is a network of filaments (White 1996) and the
filaments became longer, straighter and clumpier as the hierarchical
collapse proceeds (Summers 1996).  Clusters of galaxies, especially the
richest ones, form mostly at the vertices where several filaments
intersect.  They grow nonlinearly by gravitationally attracting matter
along the filaments and also strengthen and align the filaments at the same
time.  The dominance of filaments in the visual impression of large scale
matter distribution could be understood {\it in part} by the facts that the
filaments as mostly intersecting curves between pancakes should naturally
have higher density contrast than pancakes themselves (see West, Villumsen
\& Dekel 1991), and that a simulated universe is most often presented as a
two-dimensional projection or a slice cut where sheetlike structures cannot
be easily seen.  It also depends to some degree on choosing a right level
of density contrast.  Thus, for example, knots (clusters) will become
dominant if one chooses to look at only the highest density peaks.
 
Although formation of shocks and subsequent heating of the gas by them,
when matter accretes toward these large-scale coherent structures, are
implicit in any cosmological hydrodynamic simulations, the visual
impression of accretion shocks has became clear only after high resolution
Eulerian codes were introduced to the numerical cosmology (\eg Ryu \etal
1993; Bryan \etal 1995).  Interested readers are referred to Kang \etal
(1994b) for the details of a comparison of various cosmological
hydrodynamic codes.  They have demonstrated the accretion shocks can be
captured clearly in the simulations done by these high-resolution Eulerian
codes, while their existence is not so obvious in the simulations done by
the particle codes based on the SPH method (see their Figs. 5).  Fig. 1
shows a slice cut through a simulated universe based on the standard CDM
model previously reported by Kang \etal (1994a).  Readers are referred to
their paper for the details of the cosmological model parameters, which
should not be crucial to the current discussion.  The left panel shows the
X-ray luminosity modeled by $L_{\rm x}=\rho_{\rm b}^2 T^{1/2}$ and could be a
representative distribution of X-ray clusters.  Being weighted toward the
higher density region, this shows the knot-like distribution with some
alignments of several knots into filaments/sheets.  The right panel
shows the gas temperature distribution.  The hotter region of $T \ge
10^6$\,K is represented by solid contour curves, while the colder region of
$T< 10^6$\,K by dotted contour curves.  Shocks can be seen clearly as
strong gradients in temperature distribution, and they encompass the region
of moderately overdense regions of sheetlike/filamentary structures.  The
most prominent cluster in the lower-left corner has been shown in Fig. 1 of
KRJ96 in which the velocity pattern clearly shows accretion flows from the
background IGM and along the filaments toward the cluster.  These figures
demonstrate that the accretion shocks associated with the large-scale
coherent structures do indeed exist, even though any direct observations of
such shocks have not been made yet.  It is apparent from the diffusive
shock acceleration theory that some particles are accelerated by these
large-scale shocks.  The accretion shocks around clusters of galaxies,
being the fastest shocks generated by the deepest potential wells, are our
focus in the present paper.

\subsection{Self-Similar Evolution of Clusters}
 
The temperature of the intracluster medium (ICM) within a cluster can be
obtained with a reasonable accuracy from the X-ray observations (\eg David
\etal 1993).  Fortunately, important physical parameters such as the
velocity of an accretion shock, the velocity dispersion of galaxies, and
the depth of the potential well of the cluster can be related rather well
with the ICM temperature.  Thus one can deduce such parameters from the
observed temperature of X-ray clusters, for example, by assuming that the
gas is in a hydrostatic equilibrium with the gravitational potential due to
total mass including both the dark and baryonic matter.  Here we take
another approach in which one-dimensional (1D) spherical collapse of
clusters is modeled as the collapse of an initially overdense point-mass
perturbation followed by the secondary infall of background medium.  Such
models have been studied both semi-analytically (Fillmore \& Goldreich
1984; Bertschinger 1985) and numerically (Ryu \& Kang 1996).  In such an
accretion flow the infalling baryonic matter is stopped by a shock, while
the collisionless matter forms many caustics.  In this paper we assume that
the evolution of clusters and the properties of the accretion shocks can be
represented by such flows.  Although this 1D approach does not account for
the virialization of the central region of clusters, we will show below
essential characteristics of real clusters can be related with parameters
of the accretion shock through this model.
 
One can expect the evolution of such a flow and its accretion rate should
be dependent upon the expansion of background universe for the given
initial power spectrum of density perturbations (\ie $P(k)$ is constant for
a point-mass perturbation).  The flow solution in an $\Omega_0=1$ universe
approaches a self-similar form (Bertschinger 1985) due to the scale-free
nature of the problem and so can be treated analytically.  In a low density
universe (\ie $\Omega_0<1$), however, either open or flat with the
cosmological constant, the solution is not self-similar and can be studied
only numerically (Ryu \& Kang 1996).  Here we will consider the accretion
flows only in the Einstein-de Sitter universe ($\Omega_0=1$), and adopt the
one-dimensional (1D), self-similar accretion given by Bertschinger (1985)
as an evolutionary model for clusters and the accretion shocks.
 
Since the solution is self-similar, the shock parameters such as the radius
and velocity of the accretion shock, and the postshock gas temperature can
be uniquely determined by a single parameter at a given epoch.  For
example, the mass contained inside the outermost caustic, $M_{\rm c} =
M(r<r_{\rm c})$ at present epoch can be such a parameter.  The radius and
velocity of the accretion shock, and the postshock temperature at $r =
0.3r_{\rm s} = 0.64 h^{-1}$ Mpc are given by $r_{\rm s} = 2.12 h^{-1} {\rm
Mpc} (M_{\rm c}~h/10^{15} M_\sun)^{1/3}$, $V_{\rm s} = 1.75{\times} 10^3
\kms (M_{\rm c}~h/10^{15} M_\sun)^{1/3}$, and $T_{\rm c}(0.3r_{\rm s})=
6.06 {\rm keV} (M_{\rm c}~h/10^{15} M_\sun)^{2/3}$, respectively (Ryu \&
Kang 1996).  The reason that $T_{\rm c}(0.3r_{\rm s})$ is an interesting
quantity will be given shortly.  If one follows the evolution of a given
perturbation, on the other hand, the length scale of the accretion flow
which is proportional to the turn-around radius will grow with time as
$r_{\rm ta} \propto t^{8/9} \propto (1+z)^{-4/3}$.  The position of the
shock, $r_{\rm s}$, in units of $r_{\rm ta}$ and the velocity of the shock,
$V_{\rm s}$, in units of $r_{\rm ta}/t$ are fixed, so the shock radius and
velocity for the same perturbation evolves as $r_{\rm s} \propto
(1+z)^{-4/3}$ and $V_{\rm s} \propto (1+z)^{1/6}$.
 
Here we attempt to relate the physical parameters of the accretion shocks
with observed temperature of X-ray clusters which often represents the
emission weighted temperature in the core within $\sim 0.5 h^{-1}$ Mpc.  We
note that the temperature of the postshock gas in the 1D self-similar
solution increases toward the center, in fact, to infinity, while observed
ICM temperature distribution of most clusters is nearly isothermal, and in
some cases shows a clear depression towards the center due to cooling.
This discrepancy comes about, because the extrapolation of the self-similar
evolution to $t\to t_i$ ($r \to 0$) is not valid during the initial
collapse when the accreted mass is not much greater than the mass inside
the initial perturbation.  Also the 1D self-similar solution does not
account for the virialization of the core region.  According to the 3D
cosmological hydrodynamic simulations (\eg Crone, Evrard \& Richstone 1994;
Kang \etal 1994a; Navarro, Frenk \& White 1995), the ICM is shock heated to
the virial temperature and then settles into hydrostatic equilibrium with
an approximate isothermal structure, which is consistent with the observed
temperature distribution of the clusters.  Cosmological SPH simulations
(Navarro \etal 1995; Evrard, Metzler \& Navarro 1995) showed that simulated
clusters of different masses in fact have similar structures when scaled to
a fixed density contrast (\eg $\delta(r)= \bar\rho(r)/\rho_{\rm crit} \gsim
200{-}500$).  Navarro \etal also showed that the temperature profile of
simulated clusters can be approximated within a factor of two by that of
the 1D self-similar solutions for the outer region, $r \gsim 0.3 r_{200}$
(where $r_{200}$ is the radius at $\delta=200$), while the inner region, $r
< 0.3 r_{200}$, is isothermal.  {From} this consideration, we have made a
simple approximation that the observed X-ray temperature of a cluster is
similar to $T_{\rm c} (0.3r_{\rm s})$ of the self-similar solutions, and so
it provides the direct information about the shock velocity and radius from
the self-similar solutions.
 
{From} the equations given above which relates $r_{\rm s}$,
$T_{\rm c}(0.3r_{\rm s})$, and
$V_{\rm s}$ with the cluster mass $M_{\rm c}$ and their redshift dependences,
one can find them as a function of $kT_{\rm obs} = T_{\rm c}(0.3r_{\rm s})$
and $z$ in an $\Omega_0=1$ universe as follows.
\begin{eqnarray}
r_{\rm s} &=& 2.12 h^{-1}\,{\rm Mpc}\;
\left({kT_{\rm obs}\over 6.06{\rm keV}}\right)^{1/2}\;(1+z)^{-3/2}\\
V_{\rm s} &=& 1.75{\times} 10^3\,\kms
\left({kT_{\rm obs}\over 6.06{\rm keV}}\right)^{1/2}.
\end{eqnarray}
Evrard \etal, on the other hand, showed that $r_{500}$ where
$\delta(r_{500})=500$ is a {\it conservative} estimate for the boundary
separating the inner virialized region from the outer infalling flow.
They gave the scaling relation $r_{500} = 0.965\,h^{-1}\;{\rm Mpc}\times
({kT/ 6.06\,{\rm keV}})^{1/2}$ for clusters at $z=0$ for
an $\Omega_0=1$ universe.
The fact that the ratio of the shock radius of the self-similar
solution to the characteristic radius of virialized region in 3D
simulation is $r_{\rm s}/r_{500}=2.2$ seems reasonable, since
$r_{\rm s}$ corresponds to the radial position at $\delta \sim 80$ and
so it should be larger than $r_{500}$.
According to Ryu \& Kang (1996), the clusters of a given temperature
have smaller accretion velocity by less 20\% for $\Omega_0=0.3$ in
either open or flat with $\Lambda_0\neq 0$ universes.
Since we use observed temperature distribution for the cluster
abundance to be discussed below, the main results of our model should
remain valid for a low $\Omega_0$ cosmology.
 
According to Bertschinger (1985), the gas density upstream to the shock
is $\rho_1 = 4.02 \Omega_{\rm b}\,\rho_{\rm crit}(z) =
 7.56{\times}10^{-29}h^2\,{\rm g\,cm^{-3}}\times \Omega_{\rm b} (1+z)^3$.
Thus all the shock parameters necessary for our model
can be obtained from the observed redshift and X-ray temperature. In the
following we adopt $\Omega_{\rm b} = 0.06$ in all numerical calculations.
 
\subsection{Magnetic Field and Diffusion Coefficient}
 
The strength and morphology of the intergalactic magnetic fields remain
largely unknown and observational tasks to detect them are very challenging
even with today's technology (see, for a review, Kronberg 1994).
Theoretical study on the generation of the primordial fields and their
subsequent amplification during the structure formation is also still in
its infancy. In fact it is not clear at all that there was a primordial
magnetic field.  Recent study by Kulsrud \etal (1996), however, showed that
a weak seed field can be generated at shocks and then amplified via the
protogalactic turbulence during the structure formation up to equipartition
with the turbulence. An alternative theory (Biermann 1996) uses magnetic
stellar winds as the sources of magnetic fields for galaxies and their
environment.  On observational fronts, there exist some concrete
observations that can be used in inferring the general distribution of the
magnetic fields in intergalactic space.  The current observational upper
limit for a large-scale, pervading field is about $10^{-9} \,r_0^{-1/2}
$\,gauss (Kronberg 1994), where $r_0$ is the assumed reversal scale of the
magnetic field topology in units of 1 Mpc.  Thus, if the reversal scale
throughout the universe were larger than 1 Mpc, such as $\approx 30 h^{-1}$
Mpc, the bubble-scale of the galaxy distribution, then this upper limit
would be reduced by a factor of 6.3 (for $h=0.75$).  The magnetic fields
inside typical galaxies are observed to be order of 3-10\,$\mu$gauss, and
magnetic fields near the $\mu$gauss level are also common in core regions
of rich clusters according to many recent observations (Kim, Tribble \&
Kronberg 1991; Taylor \& Perley 1993; Taylor, Barton \& Ge 1994); some
cooling flow clusters clearly show magnetic fields even higher than those
typical in the interstellar medium of galaxies (see Kronberg 1994).
Studies on the field generation via dynamos in the cooling flows (Ruzmaikin
\etal 1989) and the field inputs from radio galaxies inside clusters
(En{\ss}lin \etal 1996) have attracted some attention recently.  Here we
are concerned most with the fields near the accretion shocks on larger
scale than cluster core, that is, within ${\sim}5h^{-1}\,$Mpc around
clusters of galaxies.  Observation of such fields has been attempted by Kim
\etal (1989) in which fields of 0.1 $\mu$gauss level were deduced {\it in
the plane of the supercluster} connecting the Coma cluster and A1367 by
assuming equipartition between the magnetic fields and relativistic
particles.  This ``bridge'' of emission region of $\sim 1.125 h^{-1}$Mpc
seems to be infalling toward the Coma and is within the supergalactic plane
(so presumably inside the accretion shocks).  Vall\'ee (1990,1993), on the
other hand, suggested the existence of a magnetic field component of $\sim
1.5\,\mu$gauss in a 10\,Mpc region around the Virgo cluster, in this case
using an assumed length scale of 10 Mpc.  This might represent the fields
outside the accretion shocks encompassing the supergalactic plane.
 
Here we adopt a simple, but common assumption that the ICM magnetic
field is in rough equipartition with the thermal energy of the gas.
Then the field strength in postshock region can be estimated from the
thermal energy inside the shock (\ie in the postshock region) according to
\begin{equation}
E_{\rm th} = 1.5 {\rho_2 \over \mu m_{\rm\SSS H}} kT_2 \sim {B_2^2 \over 8\pi}
\end{equation}
where $\rho_2(z) = 4\rho_1(z)$, $\mu=0.61$, and $T_2 = 2.69{\times}
10^7\,{\rm K}\times (kT_{\rm obs}/ 6.06{\rm keV})$.  Then the postshock
field is given by
\begin{equation}
B_2= (1.71\,\mu{\rm gauss})\;f_B h \left[{\Omega_{\rm b}\over 0.06}\,
{kT\over 6.06 {\rm keV}}\right]^{1\over 2}
\Big[1+z\Big]^{3\over 2} ,
\end{equation}
where $f_B \lsim 1$ is a factor which controls the field strength in terms of
the equipartition value, that is, $f_B=1$ means the equipartition between
the field and thermal energies.  For fiducial values of $\Omega_{\rm
b}=0.06$ and $h=0.75$, this gives for $f_B=1$ the fields of
0.52-1.64\,$\mu$gauss for $kT_{\rm obs} = 1{-}10\,$keV at $z=0$.  It is
much harder to estimate the fields in the unshocked infalling flow upstream
to the shock, so here we will simply assume $B_1= B_2 (\rho_1/\rho_2)$ for
a turbulent field.  We note here that the turbulent amplification of the
field could generate some fields not only downstream but also upstream to
the shocks due to turbulent nature of the flows (Kulsrud \etal 1996).
 
We consider two kinds of models for the particle diffusion.
The theoretical minimum for the diffusion coefficient in a strongly turbulent
field with parallel geometry (\ie the mean field is parallel to the
flow direction)
is given by the Bohm formula (see, \eg, Drury 1983).
\begin{equation}
\label{Bohmk}
\kappa_{\rm B}= {r_{\rm g}\,v \over 3}  ,
\end{equation}
where $r_{\rm g}$ is the gyroradius and $v$ is the velocity of the
particle.  On the other hand, the minimum diffusion coefficient in the
perpendicular shocks (\ie the mean field is perpendicular to the flow
direction) (Jokipii 1987) is given by
\begin{equation}
\label{Jokk}
\kappa_{\rm J}= r_{\rm g}\,V_{\rm s} = 3\,(V_{\rm s}/v)\,\kappa_{\rm B}.
\end{equation}
For high energy particles, $v\sim c$, so $\kappa_{\rm J}$ is smaller
than $\kappa_{\rm B}$ by a factor of $V_{\rm s}/c$.
 
The rate of diffusive acceleration is determined by the diffusion
coefficients $\kappa = \kpar \cos^2\theta + \kappa_\perp \sin^2\theta$,
where $\kpar$ and $\kappa_\perp$ are the diffusion coefficient parallel and
perpendicular to the magnetic field, respectively, and $\theta$ is the
angle between the field and the shock normal (Jokipii 1987).  Here we will
follow Jokipii (1987) in assuming that $\kappa_\perp/\kpar = ( 1 + \eta^2 )
^{-1}$ and $\kpar = \eta r_{\rm g} v /3$, where $\eta$ is the ratio of the
mean free path parallel to the magnetic field to the gyroradius.  This is
the result from standard kinetic theory which does not take account of
field-line meandering.  The Bohm diffusion coefficient corresponds to the
case where the field is turbulent on a scale of $r_{\rm g}$ {\em at all
momenta}, so it is effectively equivalent to the case of $\eta \to 1$ and
$\kappa_\perp \sim \kpar$ with a random distribution of the obliquity.  On
the other hand, the minimum diffusion coefficient in a perpendicular shock,
which is referred as the Jokipii diffusion throughout this paper,
corresponds to the case of $\theta=90^{\circ}$ and $\eta_{\rm max} \sim
c/(3V_{s}) = 100/V_{\rm s,3}$, where $V_{\rm s,3}$ is the shock velocity in
units of $10^3\,$\kms.  The latter is derived from the condition that
$\kappa_{\perp,\rm min}\sim r_{\rm g} V_{\rm s} \sim (\eta r_{\rm g}
c)/[3(1+\eta^2)]$, which implies that the particles should be scattered
before they drift through the shock.  This condition is necessary to
maintain the isotropy of the particle distribution.

\subsection{Energy Losses and Maximum Energy}
 
The mean acceleration time scale for a particle to reach a momentum
$p$ is determined by the velocity jump at the shock, and the
diffusion coefficient (\eg, Drury 1983; Jokipii 1987), that is
\begin{equation}
\tau_{\rm acc} = {\chi\over \chi-1}\,{v\over V_{\rm s}}\,{\eta r_{\rm g}\over
V_{\rm s}}\;{\cal J}(\chi,\theta)
\end{equation}
where $\chi$ is the compression ratio of the shock and
\begin{displaymath}
{\cal J}(\chi,\theta) = 
\left[ \cos^2\theta + {\sin^2\theta \over 1 + \eta^2}\right]
+
{\cos^2\theta + \chi^2\sin^2\theta /(1 + \eta^2)\over
\left[\cos^2\theta +
        \chi^2 \sin^2\theta\right]^{3/2} }\;.
\end{displaymath}
In the strong shock limit (\ie $\chi=4$), it can be written as
\begin{equation}
\label{taccobl}
\tau_{\rm acc} = (4.23{\times} 10^8\,{\rm years}) \times 
{\eta E_{18} \over B_{\mu} V_{\rm s,3}^2}\;{\cal J}(4,\theta)\,,
\end{equation}
where $E_{18}$ is the particle energy in units of EeV, $B_{\mu}$
is the field strength in units of microgauss.

The protons lose energy due to pair production and pion production on
the CBR not only on their way to earth, but also during their
acceleration at the cluster shocks.  At energies around 10 EeV the
energy loss due to pair production is dominant and the loss time scale
can be as short as $(5{\times} 10^9{\rm years})/(1+z)^3$. Above the GZK
cutoff at about 60\,EeV photopion production becomes dominant and the
loss time scale can be as short as $(5{\times}10^7{\rm years})/(1+z)^3$.
The maximum energy $E_{\rm max}$ up to which the protons can be
accelerated by the cluster accretion shocks is found by setting
$\tau_{\rm acc} = \tau_{\rm int}$, with $\tau_{\rm int}$ being the time
scale for interaction losses with the CBR, given by
\begin{eqnarray}
\label{Emax}
\tau_{\rm int} &=& \frac{2\pi^2\hbar^3 c^2\gamma^2}{k \Theta(z)}\times\\
&&\left[\int_{\epsilon'_0}^{\infty}d\epsilon'\,\epsilon'\,
\langle\sigma\kappa\rangle(\epsilon')\;
\ln\left[1 - \exp\left({-\epsilon'\over 2\gamma k \Theta(z)}\right)\right]
\right]^{-1}\!,\nonumber
\end{eqnarray}
where $\gamma$ is the Lorentz factor of the cosmic ray protons,
$\Theta(z) = \Theta_0(1+z)$ the temperature of the cosmic
microwave background at epoch $z$, and $\langle\sigma\kappa\rangle$
the inelasticity weighted cross section of the $p\gamma$ reaction,
averaged over all final states (Rachen \& Biermann, 1993). Equation
(\ref{Emax}) considers both pair production and pion production on the
CBR; the impact of a putative cosmic infrared background is
negligible for the determination of $E_{\rm max}$.
 
\begin{figure}
\epsfysize=3.5in\epsfbox[115 15 370 255]{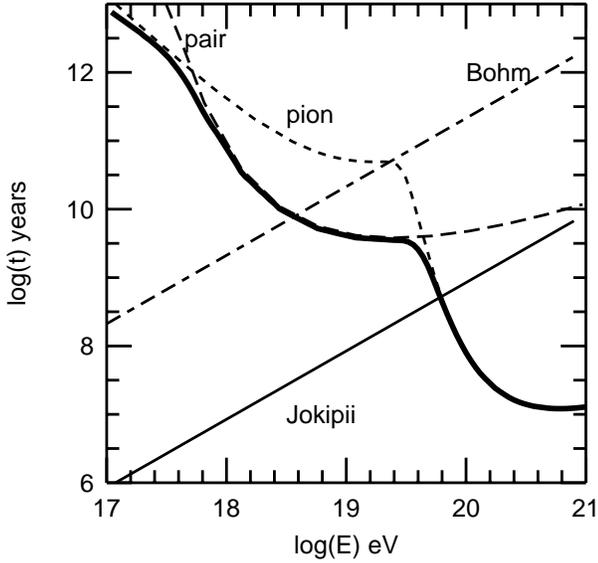}
\caption[]{Schematic representation of acceleration and energy loss time
scales as functions of particle energy.  The dotted and dashed lines
represent the loss time scale due to pion-production and pair-production,
respectively, on both CMB and infrared background radiation. The heavy
solid line is the added loss time scale due to both interactions, and it
turns out that the infrared contribution is nowhere relevant. The
dot-dashed and light solid line represent the acceleration time scales for
Bohm and Jokipii diffusion models, respectively.  For this case, adopted
shock parameters are $V_{\rm s}=1000$\kms and $B=1 \mu$gauss.  The intersection
point of both curves determines the maximum energy of acceleration.}
\end{figure}
 
In Fig. 2 we have shown the loss time scales due to the interactions with
CBR at $z=0$, and the acceleration time scales in both Bohm and Jokipii
diffusion limits.  Here the acceleration time scales for $V_{\rm s}=1000$\kms and
$B_1=1\mu$gauss are plotted.  For other values of $V_{\rm s}$ and $B$, the
acceleration time scales can be found by the following scaling relations:
$\tau_{\rm acc,B}\propto V_{\rm s}^{-2} B^{-1}$ for the Bohm limit, and
$\tau_{\rm acc,J}\propto V_{\rm s}^{-1} B^{-1}$ for the Jokipii limit.  The
maximum energy $E_{\rm max}$ for a shock of these parameters can be found
by the intersection of two curves of $\tau_{\rm int}$ and $\tau_{\rm acc}$.
The characteristic shape of loss time scale $\tau_{\rm int}$ as a function
of energy causes only a weak dependence of $E_{\rm max}$ on $\tau_{\rm
acc}$, if $\tau_{\rm acc}\lsim 3{\times} 10^{9}\,$years, but a strong
dependence if the acceleration is slower.  This time scale corresponds to
$\tau_{\rm int}$ for $E \sim 10^{19.6}$eV, Thus one can see from Fig. 2
that, for a canonical accretion shock, the maximum energy for the Bohm
limit is most likely to be set by the pair-production loss and ranges from
1 to 10\,EeV, while that for the Jokipii limit is set by the
pion-production loss at around 50\,EeV.
 
\begin{figure}
\epsfysize=3.5in\epsfbox[95 35 350 275]{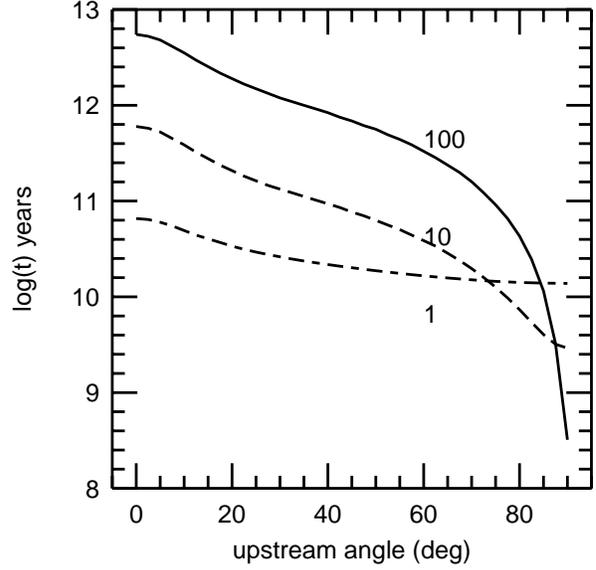}
\caption[]{Time scales for the protons to be accelerated to $E=10^{19.6}$eV
by a shock of $V_{\rm s}=1000$\kms and $B_1=1\mu$gauss as a function for obliquity.
The curves are labeled with the values of $\eta = \lambda/r_{\rm g} = 1,\,10$, and
100.}
\end{figure}
 
According to equation (\ref{taccobl}), the acceleration time scale is in
general dependent upon the obliquity and the strength of turbulent field.
Fig. 3 shows the acceleration time scales for the particles to reach up to
$E=10^{19.6}$ eV for $\eta=1, 10$ and 100 as a function of the upstream
oblique angle $\theta$.  Once again the acceleration time scales for
$V_{\rm s}=1000$\kms and for $B_1=1\mu$gauss are plotted.  This shows that
the particles can be accelerated above $10^{19.6}$eV only when the mean
field is nearly perpendicular to the shock normal.  The obliquity
dependence of $\tau_{\rm acc}$ is rather weak in the strong scattering
limit ($\eta \sim 1$), while it becomes very strong when the cross-field
diffusion is small ($\eta \sim 100$).  One can also note that $\tau_{\rm
acc}$ linearly increases with $\eta$ in quasi-parallel shocks (\ie $\theta
\approx 0^{\circ}$), so the acceleration is most efficient in the limit of
strong turbulences (\ie $\eta \to 1$).  On the other hand, $\tau_{\rm acc}$
is nearly inversely proportional to $\eta$ in quasi-perpendicular shocks
(\ie $\theta \approx 90^{\circ}$), so the acceleration becomes most
efficient in the limit of weak scattering (\ie $\eta \to \eta_{\rm max}$).
 
\subsection{Injection Spectrum}
 
For each cluster with given values of $kT_{\rm obs}$ and $z$, the
self-similar solution gives the shock parameters, $\rho_1$, $V_{\rm s}$,
$r_{\rm s}$, and $B_1$, and the time scale condition in equation
(\ref{Emax}) gives the estimate for the cutoff energy in the particle
spectrum.  Momentum distribution of the protons at the shock is assumed to
be a power law whose index is given by the parameter $\alpha$, and with an
exponential cutoff at the maximum energy ($E_{\rm max}=cp_{\rm c}$).  The
minimum momentum for the power law is $p_{\rm inj}\sim m_p V_{\rm s}$.  The
power-law index is $\alpha = 4$ in the limit of strong shocks with
dynamically insignificant CR energy density (\ie test-particle limit), but
can vary slightly around this value for real shock waves (we note that
$\alpha=4$ corresponds to a power-law in the energy spectrum of $E^{-2}$).
Here we consider values of $\alpha$ equal to and slightly larger than 4.
We further assume that the process with which the particles escape from the
shock is momentum-independent, that is, the particle spectra injected into
intergalactic (IG) space has the same shape of the proton spectra at the
shock.  Then the injected distribution from a cluster with a given ICM
temperature is given by
\begin{equation}
f(T,z,p) = A(T,z)\;p^{-\alpha}\exp\left[-{p\over p_{\rm c}(T,z)}
\right]\;\;,
\label{fsheq}
\end{equation}
where $A(T,z)$ is a normalization constant.  It is chosen by assuming that
a small fraction ($\epsilon$) of the kinetic energy density of the
infalling matter in the shock rest frame ($\rho_1V_{\rm s}^2$) is converted
to CR energy and then injected into IG space.  So the free parameter,
$\epsilon$, along with the assumed value of $\alpha$ controls the amplitude
$A(T,z)$ according to
\begin{equation}
 \epsilon=
{4\pi\over \rho_1 V_{\rm s}^2} \int_{p_{\rm inj}}^{p_{\rm c}} f(T,z,p)\,E\,p^2
\,dp\;.
\end{equation}
Thus, for given values of $\epsilon$ and $\alpha$, the normalization
constants were numerically calculated for given values of $T$ and $z$.
For $\alpha=4$, this gives an approximation for $A(T,z)$
such as
\begin{equation}
A(T,z) \simeq {{\epsilon \rho_1 V_{\rm s}^2} \over
 4\pi m_pc^2  \ln \left({p_{\rm c}/m_pc}\right)}\;\;.
\end{equation}
This along with $p_{\rm c}= (1/c)\,E_{\rm max}(T,z)$ completely specifies the
injection spectrum from a cluster.
 
In KRJ96, it was assumed that the particles escape from the shock once they
diffuse upstream to a distance comparable to the radius of the shock due to
the lateral diffusion.  Since this escape process depends on the diffusion
length, the particle spectrum injected into the IG space has a strong
dependence on the momentum, that is, only highest energy particles with
longest diffusion length can diffuse out from the shock.  We note that, if
we take the same momentum-dependent escape model, the resulting spectrum
cannot fit the observed CR distribution over a wide range of the particle
energy as well as the spectrum of our momentum-independent escape model
can.  We will also discuss the escape process in \S 3.1.

\subsection{Cluster Distribution Function}
 
The evolution of the cluster population is a critical issue which has been
under intense discussion recently, since it can provide a potentially
powerful tool for discriminating different cosmological models, especially
the matter density of the universe (\ie $\Omega_0$).  Theoretical
prediction based on a hierarchical clustering model is that clusters are
fainter but more abundant in the past (Kaiser 1986).  The evolution of
cluster population can be calculated by using various methods based on
so-called Press-Schechter formalism, if the background cosmology (\ie
$\Omega_0$ and $\Lambda_0$) and the initial density power spectrum are
given (\eg Eke, Cole \& Frenk 1996; Kitayama \& Suto 1996; Bond \& Myers
1996).  The density fluctuations of cluster mass ($M\sim 10^{15} M_{\sun}$)
would form earlier in low $\Omega_0$ universe than in high $\Omega_0$
universe, while their formation rate at the present epoch would be higher
in low bias models (\ie larger $\sigma_8$, Cen \& Ostriker 1994; Kitayama
\& Suto 1996; Bond \& Myers 1996).  Theses studies as well as most
numerical studies (Kang \etal 1994a; Tsai \& Buote 1996) indicate that the
standard CDM model of a critical density universe produces too many
clusters compared to observed local cluster abundance.  Also recent
observations of distant X-ray clusters at $z<0.3$.  (Castander \etal 1995;
Ebeling \etal 1995) seem to indicate that clusters have evolved very little
or not evolved at all in this redshift range, which is more consistent with
low bias or low density models.
 
In order to model the distribution of cosmological population of
clusters, first we adopted a temperature distribution function of
clusters given by Henry and Arnaud (1991) which was derived from the
observed local clusters.  The number density in unit comoving volume
at the present is given by for $kT=$1--10\,keV,
\begin{equation}
 n_{\rm o} (kT) = 1.8{\times} 10^{-3}\,h^3\,{\rm Mpc}^{-3}\,{\rm keV}^{-1}
\left({kT\over {\rm keV}}\right)^{-4.7}
\label{henry}
\end{equation}
Secondly, we made a simple assumption that the comoving density of cluster
evolves as a power-law of a scale factor, $(1+z)$, that is, $n(kT,z) =
n_{\rm o}(kT) (1+z)^m$.  The value of $m$ at redshifts $z<0.3$ is most
likely zero or slightly negative according to the observations mentioned
above.  According to the scaling law of Kaiser (1986) for a scale free
power spectrum in a critical density universe, for CDM like density power
spectrum of $n_{\rm eff}\sim -1$ at the cluster mass scale, the index $m$
would be $-0.7$ for the above power-law temperature distribution given by
equation (\ref{henry}).  Here we will consider a range of values for $m$
($-1\,{\le}\,m\,{\le}\,{+}1$).  We take the minimum value of the redshift
as that of the Virgo cluster, that is, $z_{\rm min}=0.0036$.  Most X-ray
bright, rich clusters would form after $z\sim 5$ and in fact clusters of
$M=10^{15}M_{\sun}$ form around $z_{\rm f}\sim 0.6$ in an $\Omega_0=1$ universe.
{From} these considerations, we set the maximum redshift to be $z_{\rm
max}=5.$ But in fact contributions from the epoch earlier than $z \sim 1$
are insignificant even for $m=1$ modeled because of the spatial dilution
($1/r^2$ factor) and the interactions with CBR for distant sources.

Then number of clusters between $z$ and $z+dz$ for a given temperature
$kT$ is given by
\begin{equation}
dN(kT,z) = n_{\rm o}(kT)\;(1+z)^m\left[{dV_{\rm c} \over dz}\right]\;dz
\end{equation}
where $dV_{\rm c}/dz=4\pi (c/H_0)^3 [{1-(1+z)^{-1/2}}]^2/ (1+z)^{3/2}$ for
$\Omega_0=1$ and $(dV_{\rm c}/dz)\,dz$ is the comoving volume of a shell
defined between $z$ and $z+dz$.  Adding up the contribution from clusters
from $z_{\rm min}$ to $z_{\rm max}$ while including the interactions with
CBR, and integrating over the temperature distribution, one can get the
particle flux observed at earth according to
\begin{eqnarray}
J(E) &=& {c\over 4} \int_{ kT_{\rm min}}^{ kT_{\rm max}}\!\!d(kT) 
\int_{z_{\rm min}}^{z_{\rm max}} dz\,\\
&&\quad\left\{\left[{dN(kT,z)\over dz}\right]
\left[{ r_{\rm s}\over d_{\rm cl}(z)}\right]^2
M(E)\,f(T,z,E)\right\}\;.\nonumber
\end{eqnarray}
$d_{\rm cl}(z) = (6\,h^{-1}{\rm Gpc})\,(1+z-\sqrt{1+z})$ is the luminosity
distance of a cluster at redshift $z$, given in an $\Omega_0=1$
cosmology. $M(E)$ is the modification factor that accounts for the
interactions with CBR along the pathway to earth, calculated as defined in
Rachen and Biermann (1993), using the the continuous energy loss
approximation introduced by Berezinsky and Grigor'eva (1988). It has been
pointed out that the characteristic spikes occurring in $M(E)$ close to the
cutoff are exaggerated because of the continuous loss approximation
(Yoshida \& Teshima 1993; Protheroe \& Johnson 1996). In the integration
over redshift, however, those features are smoothed out, and it should only
be noted that the cutoff is not quite as sharp as proposed by this method.

\section{Results and Discussion}
 
\begin{figure}
\epsfysize=8.5in\epsfbox[70 56 300 610]{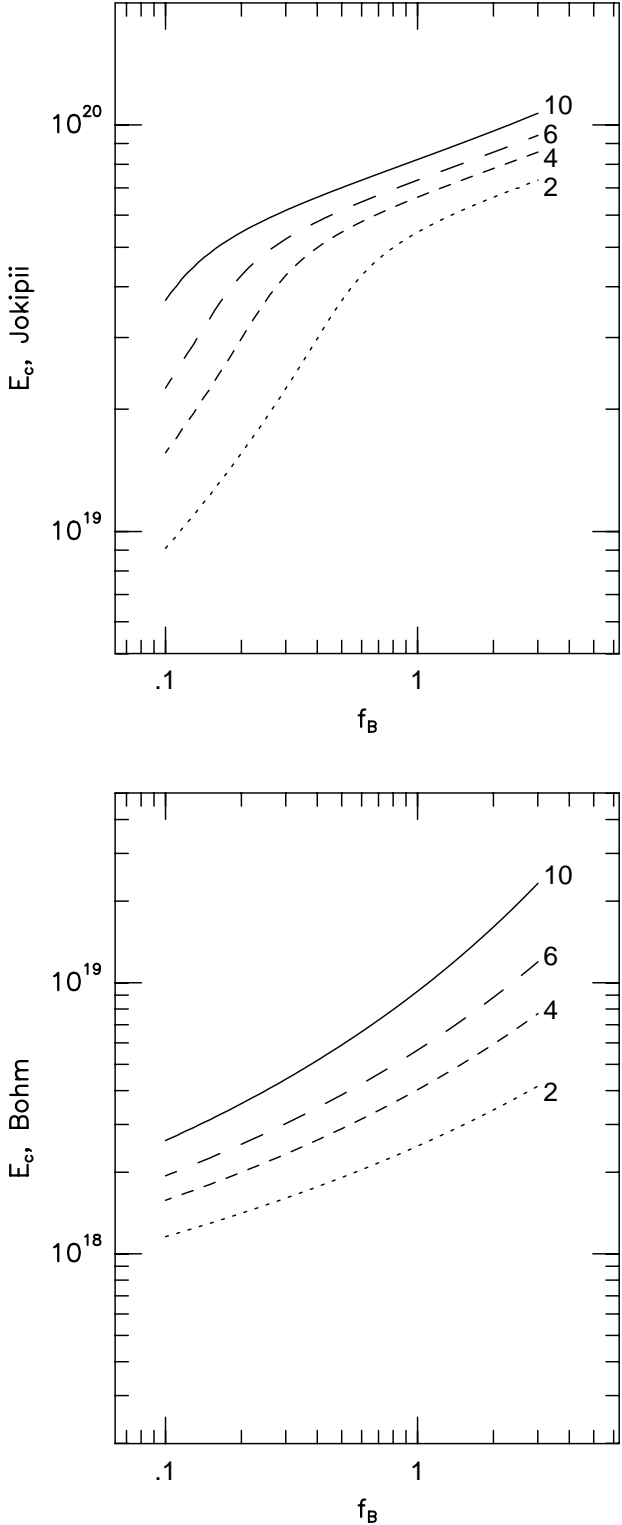}
\caption{Maximum energy of the protons accelerated by the accretion
shocks associated with the clusters with ICM temperatures of
$kT=2,\,4,\,6,$ and 10 keV at $z=0$ as a function of the parameter $f_B$
($\Omega_{\rm b} = 0.06$ and $h=0.75$).  The top panel shows models with
Jokipii diffusion, the bottom panel shows models with Bohm
diffusion.  }
\end{figure}
 
Fig. 4 shows the maximum energy calculated according to equation $\tau_{\rm
acc}=\tau_{\rm int}$ for both diffusion models for clusters with
$kT=2,\,4,\,6,$ and 10 keV at $z=0.0$ as a function of the parameter $f_B$.
As found in KRJ96, the maximum energy cannot go above the GZK-cutoff for
the Bohm diffusion coefficient.  Thus if the general field direction is
radial in the accretion flow, then the diffusive acceleration is too slow
to accelerate the protons above the GZK-cutoff. We note from Fig.~3,
however, that the transition from quasi-perpendicular Jokipii diffusion to
Bohm diffusion does not lead to a linear decrease of $E_{\rm max}$, so that
the model can work also for moderately oblique shocks with $\eta \gg 1$.

Fig. 5 shows the expected proton spectrum from the cluster accretion shocks
calculated for both diffusion models.  The power-law index for the
cosmological evolution is $m=0$, the magnetic field strength parameter is
$f_B=1$ for all models.  Different values of the injection energy fraction
$\epsilon$ is assumed for each value of the spectral index $\alpha$ to
obtain the better fit with the observation around 10\,EeV for Jokipii
models and around 1\,EeV for Bohm models, respectively.  As expected,
higher $\epsilon$ is required for steeper spectra.  For Jokipii
diffusion, the models with $4.0\,{\le}\,\alpha\,{\le}\,4.2$ show good fits
to the data, while the models with Bohm diffusion produce too few
particles above 1\,EeV.
 
The sensitivity to the cluster evolutionary model is shown in the top panel
of Fig. 6.  For all models here $\alpha=4, \epsilon=5{\times}10^{-5}$, and
$f_B=1$.  The power-law index for the evolution with the redshift for the
comoving density of clusters are $m= -1, 0,$ and $+1$ for the dotted,
solid, and dashed lines, respectively.  Since the nearby clusters
contribute most, the resulting spectrum is not severely dependent on the
evolutionary model at high redshifts.  The relatively weak cosmological
evolution of galaxy clusters compared to radio galaxies provides an even
better fit to the light component data derived from the Fly's Eye air
shower analysis (Rachen \etal 1993), but we point out that the errors are
large here and that these data have not been confirmed by the AGASA
collaboration.
 
The bottom panel of Fig. 6 shows how the results depend on the magnetic
field strength.  For all models here $\alpha=4,\;
\epsilon=5{\times}10^{-5}$, and $m=0$ is used.  The field strength relative
to the equipartition strength is represented by $f_B=0.1, 0.5$ and 1 for
the dashed, dotted, and solid lines.  One can see that cluster accretion
shock cannot produce enough particles above 10\,EeV, if the field strength
is much smaller than the equipartition value (\ie $f_B
\lsim 0.1$).  The reduction of the field from the equipartition by a factor
of up to two, however, still can give a reasonable fit to the observations.

\begin{figure}
\epsfysize=7.7in\epsfbox[56 56 300 610]{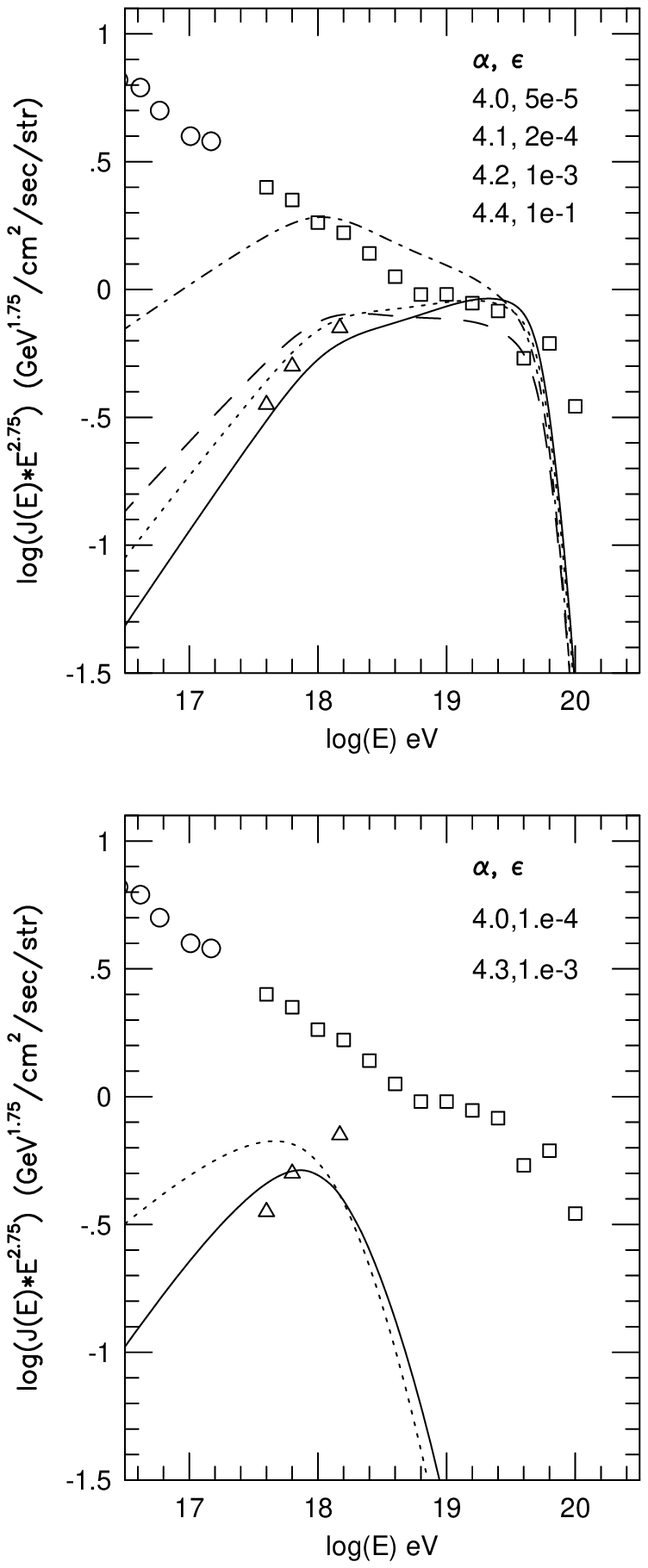}
\caption[]{Estimated proton flux from a cosmological ensemble of cluster
accretion shocks. The top panel shows the models with Jokipii diffusion.
The solid line is for $\alpha\,{=}\,4.0,\epsilon\,{=}\,5{\times}10^{-5}$,
dotted line for $\alpha\,{=}\,4.1,\,\epsilon\,{=}\,2{\times} 10^{-4}$
dashed line for $\alpha\,{=}\,4.2,\,\epsilon\,{=}\,10^{-3}$ and dot-dashed
line for $\alpha\,{=}\,4.4,\,\epsilon\,{=}\,10^{-1}$.  The bottom panel
shows the results for Bohm diffusion.  The solid line is for
$\alpha\,{=}\,4.0,\,\epsilon\,{=}\,10^{-4}$, and dotted line for
$\alpha\,{=}\,4.3,\,\epsilon\,{=}\,10^{-3}$. In all models we set $f_B =
1$, $m = 0$, $\Omega_{\rm b}=0.06$ and $h=0.75$.  The data points represent
the all-experiment data set collected by T.~Stanev (squares, see Rachen,
Stanev \& Biermann 1993), and the light component inferred from the Fly's
Eye composition measurements (triangles, ibid).}
\end{figure}
 
\begin{figure}
\epsfysize=7.7in\epsfbox[56 56 300 610]{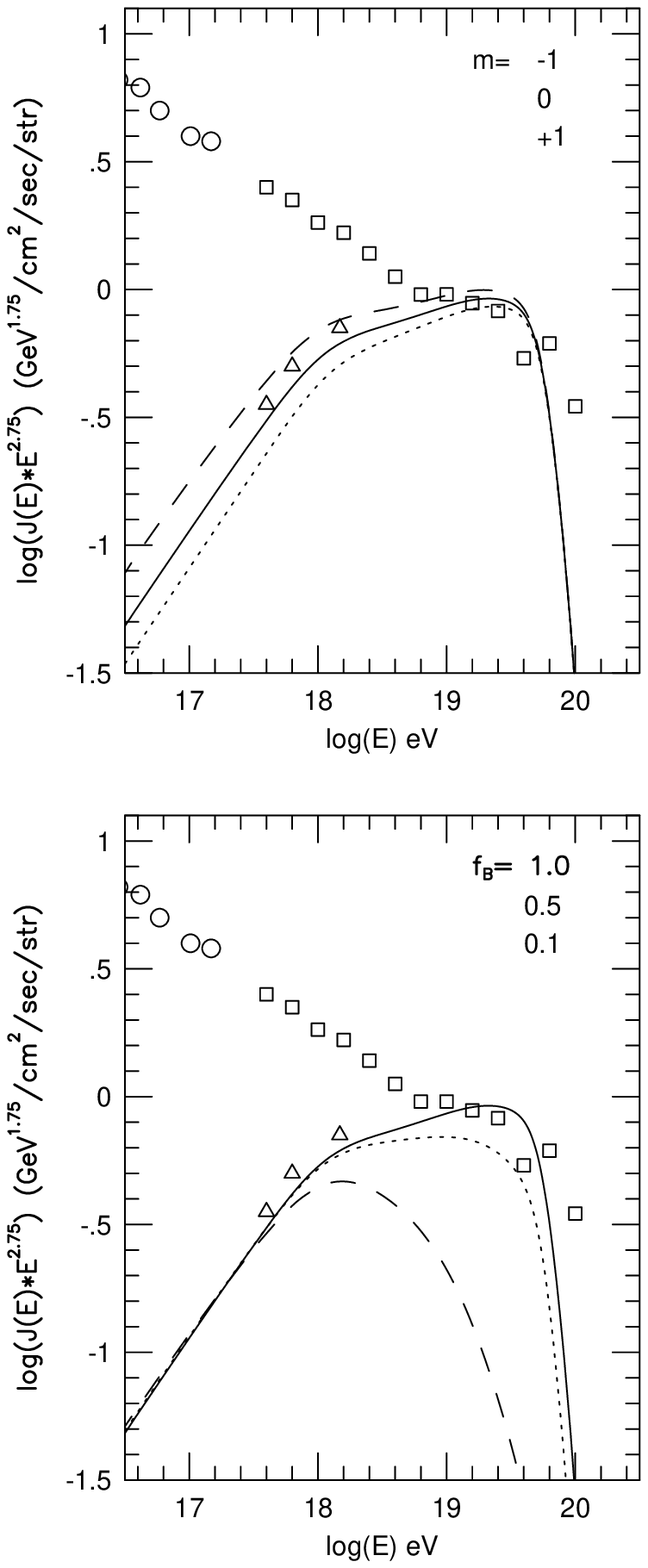}
\caption{Estimated proton flux from a cosmological ensemble of cluster
accretion shocks.  For all models here Jokipii diffusion is assumed and
$\epsilon=5 {\times} 10^{-5}$, $\alpha = 4.0$, $\Omega_{\rm b} = 0.06$ and
$h=0.75$.  The top panel shows the models with $f_B=1$, but with different
values of $m= -1,0,+1$ (dotted, solid, and dashed lines, respectively).
The bottom panel shows the models with $m=0$, but with different values of
$f_B=0.1, 0.5, 1.0$ (dashed, dotted, and solid lines respectively).  The
data points are the same as in Fig. 5.}
\end{figure}
 
\subsection{The escape process}
 
We need to consider why nature would choose an injection efficiency at a
level of order $10^{-4}$.  It is believed that a supernova blast wave can
transfer a few ($1{-}10)$\% of the explosion energy into the cosmic ray
component and the particles accelerated by the remnant shock are injected
into the interstellar medium when the shock either decays into a sonic wave
or breaks up due to various instabilities.  It would be natural to assume
that a similar fraction of the kinetic energy of the infalling flow is
likely to be transferred to the CRs.  Unlike a supernova blast wave which
expands from a point-like explosion into the interstellar space, however,
an accretion shock forms due to the converging flow associated with
gravitational collapse.  The particles will be mostly confined near the
shock and so the escape of particles by swimming against infalling flows
upstream will be very rare.  This calls for some physical models of
escape of CRs from the cluster shocks which can explain that only a
few percent of accelerated particles (which contains only a few percent of
the infall energy) is injected into the intergalactic space.  Here we must
note that the escape process cannot be energy dependent, because then the
spectrum could not match the observations.  In case of galactic cosmic rays
which travel against the solar wind to reach the earth, strong attenuation
is severely energy dependent, and so the solar wind does not provide a good
example apparently.
 
For clusters which are connected with filaments or supergalactic planes,
the particles can escape from the clusters along the filaments/sheets since
the field lines are mostly likely parallel along those structures according
to the flow velocity inside the filaments/sheets (see, for example, Fig. 1
of KRJ96).  In such a picture the low apparent efficiency is simply a
measure of the difficulty of feeding the particles into the sheets from the
surrounding of a cluster.  If this conjecture is right, the arrival
direction of UHECRs should show a correlation with the Supergalactic plane,
and so identifying a source cluster with a particular CR event may become
less stringent.  The particles can also leak when the accretion shock
becomes unstable and breaks up temporarily due to strong activities of AGNs
inside the cluster.  This could be the only possible escape route for the
particles accelerated by isolated clusters which are not part of
superclusters.  The issue is closely related with the topology and the
turbulent nature of the magnetic fields in the sheets, filaments and around
clusters and will be discussed in upcoming papers.

\subsection{The magnetic fields outside clusters}
 
For our model to work, the magnetic field just outside clusters of galaxies
must be of order 0.1 microgauss, but cannot be less, since then the Larmor
radius condition would fail, \ie the particles could not be contained.
Does this imply that we require a primordial magnetic field?
 
We note that in the evolution of clusters, repeated formation of giant radio
galaxies is likely, and they can break out and pollute the environment over
very large distances with magnetic fields.  Therefore, with such a model
tracing the large scale magnetic fields to very large radio galaxies, one would
not need a primordial magnetic field.
 
Of course, a primordial magnetic field would have to be weaker than the
observational constraints, given by the cosmological limits on the Faraday
rotation of distant radio sources; a primordial magnetic field is likely nearly
homogeneous inside the bubble structure of the galaxy distribution, and so an
upper limit of $\approx 200$\,picogauss would obtain, with a substantial
strengthening towards the spatial environment near to clusters.  If we use an
inverse Parker wind model (as an analogy to the expansion of the gas from a
radio galaxy) to do such an extrapolation, \ie scale the magnetic field
strength inversely with the distance to a cluster, then such an upper
limit is just about compatible with the magnetic fields required by our model.
 
In conclusion, we cannot decide between the two possibilities, a primordial
magnetic field, and one that derives from radio galaxies, or some other
sources such as normal galaxies, even on the large scale.

\subsection{Cosmic rays inside clusters}
 
The accretion shock produces cosmic rays scaling with the overall
energetics of the cluster, but with only some fraction of the shock
kinetic energy; thus the energy density of cosmic rays inside clusters
is expected to be less than 10\% of the thermal energy content in our
model of the intracluster medium.  On the other hand, the cosmic rays
injected from radio galaxies may produce magnetic field strengths and
energy densities of cosmic rays close to equipartition:  En{\ss}lin \etal 
(1996) proposed that radio galaxies can push the cosmic ray  content 
in clusters to the stability limit such as is believed to  happen in the
interstellar medium due to supernova explosions  (see, \eg Parker 1969 
for a review).  As demonstrated by En{\ss}lin \etal (1996) these two  
alternatives may be decidable in the near future with gamma-ray  
observations due to the pion decay production in pp-collisions.
 
\subsection{Arrival directions of UHECRs}
 
There is a study of the Haverah Park events by Stanev \etal (1995), which
shows that the arrival directions of the UHECRs are not uniform, but seem
to show positive correlations with the supergalactic plane; a related study
of the AGASA events by Hayashida \etal (1996) supports a connection between
the supergalactic plane and the arrival directions of at least a
significant fraction of UHE CRs.  This supports models for the
extragalactic origin of the highest energy cosmic rays, especially the ones
in which the sources are likely to be associated with the supergalactic
plane.  They include radio galaxies and cluster accretion shocks.  We have
looked at the distribution of local clusters ($z<0.3$) and generated a map
of the cluster accretion shocks weighted by the particle flux above a few
times $10$ EeV (Kang, Rachen \& Biermann 1996).  The map shows the similar
degree of correlations with the supergalactic plane as the arrival
directions of the UHECRs found in Stanev \etal (1995) and Hayashida \etal
(1996).  This is consistent with the fact that rich clusters beyond the
local supergalactic plane ($d> 30h^{-1}$Mpc) such as the Coma (8.3keV),
Perseus (6.3keV), 3C129 (6.2keV), A3571 (7.6keV), and Centaurus (3.9 keV)
clusters are at low supergalactic latitudes.  These rich clusters are
dominant contributors of UHE CRs according to our model.  The Virgo cluster
is the most prominent cluster in the Local Supercluster, but its
temperature is only only 2.4 keV. Thus its contribution through the
accretion shock is important only for energies below GZK cutoff; of course,
it may have an additional contribution from its radio galaxies.

\subsection{The most energetic events}
 
Concerning the events above 100\,EeV, the present model can give a
straightforward explanation for the enhancement of highest energy events
detected by the Haverah Park experiment at high north galactic latitudes
(see Stanev \etal 1995, and references therein), since the Virgo and Coma
clusters are in that direction.  On the other hand, the applicability of
the hot spot acceleration model to the center-brightened radio galaxies M87
harbored by the Virgo cluster is not yet clear.  Whether the 320\,EeV Fly's
Eye event may be explained in the context of cluster accretion shock
acceleration, remains to be investigated.  The AWM7 cluster ($T=4.0$\,keV,
$z=0.0176$) might be the closest rich cluster to the general direction of
this event. Alternatively, there is a candidate radio galaxy, 3C134; its
redshift is unknown due to obscuration, but its distance can be estimated
from radio size-luminosity relations to be in the range $30{-}300\,$Mpc,
thus it may well be the closest FR-II radio galaxy (Rachen 1995). This
estimate has been recently confirmed by an infrared detection of the
central galaxy (Hartmann 1996). All in all, since the maximum energy of
acceleration may vary strongly between individual objects in both models,
one may have to consider both possibilities in the explanation of the
highest energy events.
 
\section{Conclusions and Outlook}
 
The studies on the expected CR spectrum and distribution from galaxy
cluster accretion shocks seem to give strong evidences for a significant 
UHECR contribution, which could reach up to the highest observed energies. 
Some aspects of the model need further detailed considerations, however, 
since it has to rely on several somewhat speculative assumptions. 
A first question could be if one can prove observationally the
existence of large-scale accretion shocks around clusters.  Although the
shocks and the hot component of IGM can be clearly seen to exist in most of
cosmological hydrodynamic simulations, independent of the details of
cosmological models; it is observationally challenging to detect such
shocks because the hot gas does not radiate or absorb photons at a
detectable level.  But the hot gas of $T=10^5{-}10^6$\,K heated by the
large-scale accretion shocks seems to be a major component of the
intergalactic medium in the simulated universes (Nath \& Biermann 1993;
Ostriker \& Cen 1996).  One can also ask if the model would work for lower
density universes such as open CDM models and flat CDM models with
cosmological constant, because it is expected that the accretion flows are
weaker in such cosmologies.

Another crucial assumption for our model is the magnetic field 
in equipartition with the thermal energy in the vicinity of accretion shocks.
The general topology and turbulent nature of the large-scale magnetic field
are also closely related with our particle diffusion model and the escape 
process.   
We note that the Jokipii expression for the possible range of the
diffusion coefficient in an oblique shock geometry has been derived going to
the absolute limit of what is geometrically possible; it may be an extreme
overstatement of the limit relevant in nature.  Therefore, we emphasize that 
we use this as an assumption going into the modelling.  
On the other hand, considering the loss time scale and comparing the two limits, 
Bohm and Jokipii (Fig. 2), the maximum energy possible is only moderately 
changed, even when we are 1/3 of the way between Jokipii and Bohm limits. 
Therefore, we conclude that the speculative nature of the Jokipii 
{\it limit} has relatively little influence on our essential result, 
as long as nature allows us to go two thirds of the way towards Jokipii's limit.
This demonstrates that
the theoretical studies on how a magnetic field is
generated and amplified during the structure and galaxy formation (\eg
Kulsrud \etal 1996; Ryu, Kang \& Biermann 1996; Biermann 1996), as well as
the observational studies on how the real magnetic field in the IGM is
distributed are very important to further development of our model.
 
\section*{ACKNOWLEDGEMENTS}
We thank D. Ryu for helpful discussion on the self-similar evolution
of clusters and comments on the manuscript.
We are grateful to L. Drury, the referee, for critical review of our paper.

\end{document}